\documentclass{article}
\usepackage{psfig}
\begin{document}
\title{\vspace{-3cm}
\LARGE\bf Final steps towards a proof of the Riemann hypothesis}
\author{Carlos Castro$^1$, Jorge Mahecha$^2$ \\
{\small\em $^1$Center for Theoretical Studies of Physical Systems,}\\
{\small\em Clark Atlanta University, Atlanta, Georgia, USA}\\
\smallskip
{\small\em $^2$Institute of Physics, University of Antioquia,
Medell\'{\i}n, Colombia}}
\date{\today}

\maketitle

\begin{abstract}

A proof of the Riemann's hypothesis (RH) about the non-trivial zeros of
the Riemann zeta-function is presented. It is based on the construction
of an infinite family of operators $D^{(k,l)}$ in one dimension, and
their respective eigenfunctions $\psi_s (t)$, parameterized by continuous
real indexes $k$ and $l$. Orthogonality of the eigenfunctions is
connected to the zeros of the Riemann zeta-function. Due to the
fundamental Gauss-Jacobi relation and the Riemann fundamental relation $
Z (s') = Z (1-s') $, one can show that there is a direct concatenation
among the following symmetries, $ t $ goes to $ 1/t $, $ s $ goes to $
\beta - s $ ($\beta$ a real), and $ s' $ goes to $ 1 - s' $, which
establishes a one-to-one correspondence between the label $ s $ of one
orthogonal state to a unique vacuum state, and a zero $ s' $ of the
$\zeta$. It is shown that the RH is a direct consequence of these
symmetries, by arguing in particular that an exclusion of a continuum of
the zeros of the Riemann zeta function results in the discrete set of the
zeros located at the points $s_n = 1/2 + i \lambda_n$ in the complex
plane.

\end{abstract}

\section{\bf Introduction}
\label{sec:intro}

Riemann's outstanding hypothesis (RH) stating that the non-trivial complex
zeros of the zeta-function $\zeta(s)$ must be of the form $s_n = 1/2 \pm
i\lambda_n$, remains one of the more important open problems in pure
mathematics. The zeta-function is related to the number of primes less
than a given number, and the zeros of the zeta-function have a deep
connection with the distribution of primes \cite{riemann}. References
\cite{karatsuba,patterson,titchmarsh}) are devoted to the mathematical
properties of the zeta-function.

The RH has also been studied from the point of view of physics (e.g.,
\cite{katz,berry,selvam,connes}). For example, the spectral properties of
the $\lambda_n$'s are associated with the random statistical fluctuations
of the energy levels (quantum chaos) of a classical chaotic system
\cite{main}. Montgomery \cite{montgomery} has shown that the two-level
correlation function of the distribution of the $\lambda_n$'s coincides
with the expression obtained by Dyson with the help of random matrices
corresponding to a Gaussian unitary ensemble. Planat \cite{planat} has
found a link between RH and the called $1/f$ noise. Wu and Sprung
\cite{wusprung} have numerically shown that the lower lying non-trivial
zeros can be related to the eigenvalues of a Hamiltonian having a fractal
structure. Since the literature on the topic is rather extensive we refer
the reader to a nice review of zeta-related papers which can be found in
Ref. \cite{watkins}.

Recently Pitk\"anen \cite{pitkanen} proposed a method of proving the
Riemann hypothesis based on the orthogonality relations between
eigenfunctions of a non-Hermitian operator used in super-conformal
transformations. The states orthogonal to a ``vacuum'' state correspond to
the zeros of the Riemann zeta-function. According to his proposal, the
proof of RH rests on proving the Hermiticity of the inner product and an
assumption about the conformal gauge invariance in the subspace of the
states corresponding to the zeros of the $\zeta$-function, the plausible
role of (super) conformal invariance was proposed in \cite{cc}.

In previous works \cite{cc,cj,acj} we have already explored some possible
strategies which could lead to a solution of the problem. Now we will
pursue one of them in detail. It is based on the idea already described
above (\cite{pitkanen}, see also \cite{sp1,elimorzer}) which relates the
non-trivial zeros of the $\zeta$-function and orthogonality of
eigenfunctions of the appropriately chosen operator. We are not assuming
any ad-hoc symmetries like conformal invariance, but in fact, we show why
the $ t \rightarrow 1/t$ and $ s \rightarrow \beta -s $ symmetries are in
direct correlation with the $ s' \rightarrow 1 - s'$ of the Riemann's
fundamental identity $ Z (s') = Z (1 - s') $. This is the clue to the
proof of the RH. The function $Z$ (the fundamental Riemann function) is
defined as follows
\cite{karatsuba},
\begin{equation}
Z(s)\equiv\pi^{-s/2}\Gamma\left(\frac{s}{2}\right)\zeta(s).
\label{eq:RieFund}
\end{equation}

\section{\bf Nontrivial $\zeta$'s zeros as an orthogonality relation}
\label{sec:zeros}

Our proposal is based on finding the appropriate operator $D_1$
\begin{equation}
D_1 = - \frac{d}{d\ln t} + \frac{d V}{d\ln t} + k, \label{eq:opD}
\end{equation}
such that its eigenvalues $s$ are complex-valued, and its eigenfunctions
are given by
\begin{equation}
\psi_s (t) = t^{-s+k} e^{V(t)}.
\label{eq:psi}
\end{equation}
$D_1 $ is not self-adjoint since its eigenvalues are complex valued
numbers $s$. We also define the operator dual to $D_1$ as follows,
\begin{equation}
D_2 = \frac{d}{d\ln t} + \frac{d V}{d\ln t} + k, \label{eq:opD1}
\end{equation}
that is related to $ D_1 $ by the substitution $ t \rightarrow 1/t $ and
by noticing that
$$
\frac{dV(1/t)}{d\ln(1/t)} = -\frac{dV(1/t)}{d\ln t},
$$
where $V(1/t)$ is not equal to $V(t)$.

Since $V(t)$ can be chosen arbitrarily, we choose it to be related to the
Bernoulli string spectral counting function, given by the Jacobi theta
series,
\begin{equation}
e^{2 V(t)} = \sum\limits_{n=-\infty}^\infty e^{-\pi n^2 t^l} =
2\omega(t^l) + 1.
\label{eq:JacobiTheta}
\end{equation}
This choice is justified in part by the fact that Jacobi's theta series $
\omega $ has a deep connection to the integral representations of the
Riemann zeta-function \cite{biane}.

Latter arguments will rely also on the following related function defined
by Gauss,
\begin{equation}
G(1/x) = \sum\limits_{n=-\infty}^\infty e^{-\pi n^2/x} = 2\omega(1/x)+1,
\label{eq:GaussTheta}
\end{equation}
where $ \omega(x) = \sum_{n=1}^\infty e^{-\pi n^2 x} $. Then, our $ V $ is
such that $ e^{2 V(t)} = G(t^l) $. We defined $ x $ as $ t^l $. We call $
G(x) $ the Gauss-Jacobi theta series (GJ).

Thus we have to consider a family of $D_1$ operators, each characterized
by two real numbers $k$ and $l$ which can be chosen arbitrarily. The
measure of integration $d\ln t$ is scale invariant. Let us mention that
$D_1$ is also invariant under scale transformations of $t$ and $F=e^V$
since $dV/(d\ln t)=d\ln F/(d\ln t)$. In \cite{pitkanen} only one operator
$D_1$ is introduced with the number $k=0$ and a different (from ours)
definition of $F$.

We define the inner product as follows,
\begin{equation} 
\langle f\vert g\rangle=\int\limits_0^\infty f^* g\frac{dt}{t}. 
\label{eq:innerpr} 
\end{equation} 
Based on this definition the inner product of two eigenfunctions of $D_1$
is
\begin{equation}
\begin{array}{rl}
\displaystyle\langle\psi_{s_1}\vert\psi_{s_2}\rangle
&\displaystyle=\alpha\int\limits_0^\infty
e^{2V}t^{-s_{12}+2k-1}dt\\
&\displaystyle=\frac{2\alpha}{l}
Z\left[\frac{2}{l}(2k-s_{12})\right],
\label{eq:innerprpsi}
\end{array}
\end{equation}
where we have denoted
$$
s_{12}=s_1^*+s_2=x_1+x_2+i(y_2-y_1),
$$
used the expressions (\ref{eq:JacobiTheta}) and (\ref{eq:RieFund}) and
noticed that
$$
\langle s_1\vert s_2\rangle = \langle 1/2 + i0\vert s_{12} - 1/2\rangle.
$$
Thus, the inner product of $\psi_{s_1}$ and $\psi_{s_2}$ is equivalent to
the inner product of $\psi_{s_o}$ and $\psi_s$, where $s_o=1/2+i 0$ and
$s=s_{12}-1/2$. Constant $\alpha$ is to be appropriately chosen so that
the inner product in the critical domain is semi-positively definite. The
integral is evaluated by introducing a change of variables $t^l=x$ (which
gives $dt/t=(1/l)dx/x$) and using the result provided by the equation
(\ref{eq:GaussTheta}), given in Karatsuba and Voronin's book
\cite{karatsuba}. Function $Z$ in (\ref{eq:RieFund}) can be expressed
in terms of the Jacobi theta series, $\omega(x)$ defined by
(\ref{eq:JacobiTheta}) (see \cite{patterson}),
\begin{equation}
\begin{array}{rl}
\displaystyle\int\limits_0^\infty \sum\limits_{n=1}^\infty
e^{-\pi n^2x}x^{s/2-1}dx& = \\
&\\
&\displaystyle = \int_0^\infty x^{s/2 - 1} \omega (x) dx\\
&\\
&\displaystyle ={1\over s(s-1)}+\int_1^\infty [x^{s/2-1} +
x^{(1-s)/2 - 1}] \omega (x) dx\\
&\\
&\displaystyle = Z(s) = Z(1-s).
\label{eq:ThetaRie}
\end{array}
\end{equation}
Since the right-hand side of (\ref{eq:ThetaRie}) is defined for all $s$
this expression gives the analytic continuation of the function $Z(s)$ to
the entire complex $s$-plane \cite{patterson}. In this sense the fourth
``$=$'' in (\ref{eq:ThetaRie}) is not a genuine equality. Such an analytic
continuation transforms this expression into the inner product, defined by
(\ref{eq:innerprpsi}).

A recently published report by Elizalde, Moretti and Zerbini
\cite{elimorzer} (containing comments about the first version of our
paper \cite{acj}) considers in detail the consequences of the analytic
continuation implied by equation (\ref{eq:ThetaRie}). One of the
consequences is that equation (\ref{eq:innerprpsi}) loses the meaning of
being a scalar product. Arguments by Elizalde {\it et al.\/}
\cite{elimorzer} show that the construction of a genuine inner product is
impossible.

Therefore from now on we will loosely speak of a ``scalar product''
realizing that we do not have a scalar product as such. The crucial
problem is whether there are zeros outside the critical line (but still
inside the critical strip) and not the interpretation of equation
(\ref{eq:innerprpsi}) as a genuine inner product. Despite this, we still
rather loosely refer to this mapping as a scalar product. The states still
have a real norm squared, which however need not to be positive-definite.

Here we must emphasize that our arguments do not rely on the validity of
the zeta-function regularization procedure \cite{elizalde}, which
precludes a rigorous interpretation of the right hand side of
(\ref{eq:ThetaRie}) as a scalar product. Instead, we can simply replace
the expression ``scalar product of $\psi_{s_1}$ and $\psi_{s_2}$'' by the
map $S$ of complex numbers defined as
\begin{equation}
\begin{array}{rccl}
S:& {\cal C}\otimes{\cal C}&\to&{\cal C}\\
 &&&\\
&(s_1,s_2) &\mapsto&\displaystyle
S(s_1,s_2)=-Z (as + b).
\end{array}
\label{eq:map}
\end{equation}
where $ s = s^*_1 + s_2 - 1/2$ and $ a = -2/l; b = (4k-1)/l$. In other
words, our arguments do not rely on an evaluation of the integral
$\langle\psi_{s_1}\vert\psi_{s_2}\rangle$, but only on the mapping
$S(s_1,s_2)$, defined as the finite part of the integral
(\ref{eq:innerprpsi}). The kernel of the map $S(s_1, s_2) = -Z(as + b)$
is given by the values of $s$ such that $Z(as + b) = 0$, where
$
\langle s_1\vert s_2\rangle = \langle s_o\vert s \rangle
$
and
$
s_o = 1/2 + i0
$.
Notice that $2b+a = 4(2k-1)/l$. We only need to study the
``orthogonality'' (and symmetry) conditions with respect to the ``vacuum''
state $s_o$ to prove the RH from our theorem 2. By symmetries of the
``orthogonal'' states to the ``vacuum'' we mean always the symmetries of
the kernel of the $S$ map.

The ``inner'' products are trivially divergent due to the contribution of
the $ n = 0$ term of the GJ theta series in the integral
(\ref{eq:innerprpsi}). From now on, we denote for ``inner'' product in
(\ref{eq:innerprpsi}) and (\ref{eq:map}) as the finite part of the
integrals by simply removing the trivial infinity. We shall see in the
next section, that this ``additive'' regularization is in fact compatible
with the symmetries of the problem.

\section{\bf Three theorems and a proof of the RH}
\label{sec:proof}

In our approach, the RH emerges as a consequence of the symmetries of the
orthogonal states to the ``vacuum'' state $\psi_{s_o}$. To this end we
prove now the first theorem:

{\it Th. 1\/}. If $a$ and $b$ are such that $2b+a=1$, the symmetries of
all the states $\psi_s$ orthogonal to the ``vacuum'' state are preserved
by any map $S$ (equation \ref{eq:map}), which leads to $Z(as+b)$

{\it Proof}: If the state associated with the complex number
$s=x+iy$ is orthogonal to the ``vacuum'' state and the ``scalar
product'' is given by
$
Z(a s + b) = Z (s')
$,
then the Riemann zeta-function has zeros at $s'=x'+iy'$, $s'^*$, $1-s'$
and $1-s'^*$.

If we equate $as+b=s'$, then $as^*+b=s'^*$. Now, $1-s'$ will be equal to
$a(1-s)+b$, and $1-s'^*$ will be equal to $a(1-s^*)+b$, if, and only if,
$2b+a=1$. Therefore, all the states $\psi_s$ orthogonal to the ``vacuum''
state, parameterized by the complex number $1/2 + i0$, will then have the
same symmetry properties with respect to the critical line as the
nontrivial zeros of zeta.

Notice that our choice of $a=-2/l$ and $b=(4k-1)/l$ is compatible with
this symmetry if $k$ and $l$ are related by $l=4(2k-1)$. Conversely, if
we assume that the orthogonal states to the ``vacuum'' state have the
same symmetries of $Z(s)$, then $a$ and $b$ must be related by $2b+a=1$.
This results in a very specific relation between $k$ and $l$, obtained
from $a+2b=1$ for $a,b$ real. It is clear that a map with arbitrary
values of $a$ and $b$ does not preserve the above symmetries.

{\it Th. 2 \/}. The RH is a direct consequence of the assumption that the
kernel of the map $Z(as + b)$ has the same symmetry properties as the
zeros of zeta. This means that the values of $s$ such that $Z(as + b) =
0$; i.e. the states ``orthogonal'' to the ``vacuum'' state $ s_o = 1/2 +
i0 $, are symmetrically distributed with respect to the critical line and
come in multiplets of four arguments $s$, $1-s$, $s^*$, $1 -s^*$.

{\it Proof:\/} Due to the analytic properties of the function $Z(as + b)
= Z (s') $ it follows from theorem 1 that such symmetry conditions are
satisfied if and only if:
$
a(k, l) + 2b(k, l) = 1
$,
implying that $l = 8k - 4$ from which in turn follows that:
$
s' = a(k, l) s + b (k, l) = a(k, l)(s - 1/2) + 1/2
$,
so their real parts satisfy:
$
x' = 1/2 + a(k, l)(x - 1/2)
$.

Hence, for a fixed value of $x$, the value of its real part $x'$ can be
continuously changed by continuously changing $(k,l)$, since $a = - 2/l$.
If we assume that the zeros form a discrete set of points, this means that
$x = 1/2$ is the only consistent value it can have. Unless $l = 0$ ($a =
\infty$) which is absurd because this yields a constant potential $V(t)$,
$x = t^l = 1$. The case $l = \infty$ ($a = 0$) is also ruled out because
the potential $V (t)$ either blows up, is zero or is ill-defined depending
on the values of $t$. From this follows that $x' = 1/2$ is the only
consistent and possible value which the real part of the zeros of zeta can
have. Therefore, RH follows directly from the latter conclusion.

Another way of rephrasing this is to say that the family of the
$D_1^{(k,l)}$ operators yields a continuous family of maps which map $x$
into $x'$. Such pairs of points $(x,x')$ have the double-reflection
symmetry $x\rightarrow 1-x$, $x'\rightarrow 1-x'$, if, and only if, $1 = a
+ 2b$ from which it follows that $1/2=a(k,l)/2+b(k,l)$. This means that
all the lines given by $x'=a(k,l)x+b(k,l)$ must have the common point in
the $x-x'$ plane given by $(1/2,1/2)$ for all the values of $(k,l)$
obeying $l=4(2k-1)$. If $\psi_s $ is orthogonal to the ``vacuum'' state
associated with the complex number $1/2+i0$ then $s'$ is a zero of the
$\zeta$, and the real parts of $s$ and $s'$ are related by
$x'=a(x-1/2)+1/2$. If one assumes a discrete set of zeros then these real
parts must be independent of $k,l$, that is independent of $a$. This can
be satisfied only if the orthogonal state has for its real part equal to
$x=1/2$, which yields $x'=1/2$, that is the RH.

{\it Th. 3 \/}. The $ s' \rightarrow 1 - s' $ symmetry of the Riemann
nontrivial zeros and the $ t \rightarrow 1/t $ symmetry of the ``inner''
products, are concatenated with the $ s \rightarrow \beta - s$ symmetry of
the ``orthogonal'' states to a ``vacuum'' state $ s_o = \beta/2 + i 0 $,
for any real $\beta$.

{\it Proof:\/} Gauss has shown that \cite{gaussjacobi},
\begin{equation}
G(1/x) = x^{1/2}\,G(x),
\label{eq:gauss}
\end{equation}
where the Jacobi series $G(x)$ is defined by equation
(\ref{eq:JacobiTheta}). (\ref{eq:gauss}) implies that one can always find
a $\beta$, such that $\psi_s (1/t) = \psi_{\beta - s} (t)$ for all values
of $ s $ if, and only if, $ 2k - \beta = l/4 $.  Due to ($k,l$) are real,
this forces $\beta$ be a real. In terms of ($a,b$) this relation becomes,
$1 = a (2\beta - 1) + b$, that when $\beta = 1$ gives the known relation $
1 = a + 2b $.

Then, invariance of the ``inner'' product under the inversion symmetry, $
t \rightarrow 1/t $ follows by adopting a standard regularization
procedure of removing the infinities, which yields the well defined
finite parts:
$
\langle\psi_{1/2 + i 0} (t)\vert\psi_s (t)\rangle =
\langle\psi_{1/2 + i 0} (1/t)\vert\psi_s (1/t)\rangle =
\langle\psi_{1/2 + i 0} (t)\vert\psi_{1 - s} (t)\rangle =
-Z(s') = -Z(s'')
$.
If this invariance under inversion holds for all values of $ s $ and due
to the fact that $ s' \not= s'' $ (except for the trivial case when $ 1
-s = s $, $ s = 1/2 $) the only consistent solution, for all values of $
s $, has to be $ s'' = 1 - s' $ due to Riemann's fundamental identity $ Z
(s') = Z (1 -s') $.

Then one can write down the integrals, after removing the infinities, in
explicit form as,
\begin{equation}
\int_0^\infty dx G (x) x^{s'/2 - 1} = 2 Z(s').
\label{eq:GtoZ}
\end{equation}
Under $ x \rightarrow 1/x $ we have,
\begin{equation}
\begin{array}{rl}
\displaystyle\int_\infty^0 d (1/x) G (1/x) (1/x)^{s'/2 - 1} &= \\ 
\displaystyle\int_0^\infty dx G (x) x^{(1-s')/2 - 1} &= 2 Z(1 -s').
\label{eq:infinity}
\end{array}
\end{equation}

Adopting an ``additive'' regularization procedure of removing the
infinities, one can see that (\ref{eq:GtoZ}) = (\ref{eq:infinity}). This
shows that the ``inner'' products are invariant under $ t $ goes to $ 1/t
$.

The origins of the symmetry $t\rightarrow 1/t$ in the scalar product
$\langle s_o\vert s\rangle$ stem from the invariance of the integral
(\ref{eq:GtoZ}) (modulo the infinities) under the $ x \rightarrow 1/x $
transformation. Such invariance is translated as an invariance under $ s'
\rightarrow 1-s' $, based on the Gauss-Jacobi relation. Notice how
important it is not to introduce {\it ad hoc\/} any symmetries, like
conformal invariance, without justifying their origins. We are basing
everything in the fundamental relation $ Z (s') = Z (1-s') $, therefore
our symmetry $t \rightarrow 1/t$ is well justified.

From the symmetries of theorem 3, one can easily show that
$ a + 2b = 1 $.
Then, the RH will follows immediately, since,
$ s' = as + b = as + (1 - a)/2 = a (s - 1/2)) + 1/2 $.
The real values of this equation are
$ x' = a (x - 1/2) + 1/2 $.

And as we have seen earlier in section \ref{sec:zeros}, because $ a = -
2/l $ depends continuously on the parameter $ l $, for any given fixed
value of $ x $, one could always assign in a continuous manner zeros whose
real parts are of the given form. This can be achieved by simply varying
in a continuous fashion the parameter $ l $. If, and only if, the zeros
are discrete the only compatible and consistent solution is,
$
x = (1/2) \Leftrightarrow x' = (1/2) \Leftrightarrow {\rm RH~is~true}
$.

The ``vacuum'' state can be defined in many ways. We can show that any
``vacuum'' state must have the form $s_o = \beta/2 + i 0$. If $f(s) =
\beta - s$, the fixed point of $f$ is such that $ \beta - s_o = s_o$,
gives $s_o = \beta/2$. The orthogonal states to the new ``vacuum'' are
such that,
$
\langle s_o\vert s\rangle = \langle 1/2 + i 0\vert s + s_o - 1/2\rangle =
-Z [a (s + s_o - 1/2) + b] = -Z (s') = 0
$.

Now we will demonstrate how by choosing a continuous family of operators
with $l = 8k -4 $ ({\it i.e.\/} $ a + 2b = 1$), the RH is a direct
consequence of the fact that the states orthogonal to the ``vacuum'' state
have the same symmetry properties as the zeros of $\zeta$-function.

\begin{figure}[h]
\centerline{\psfig{figure=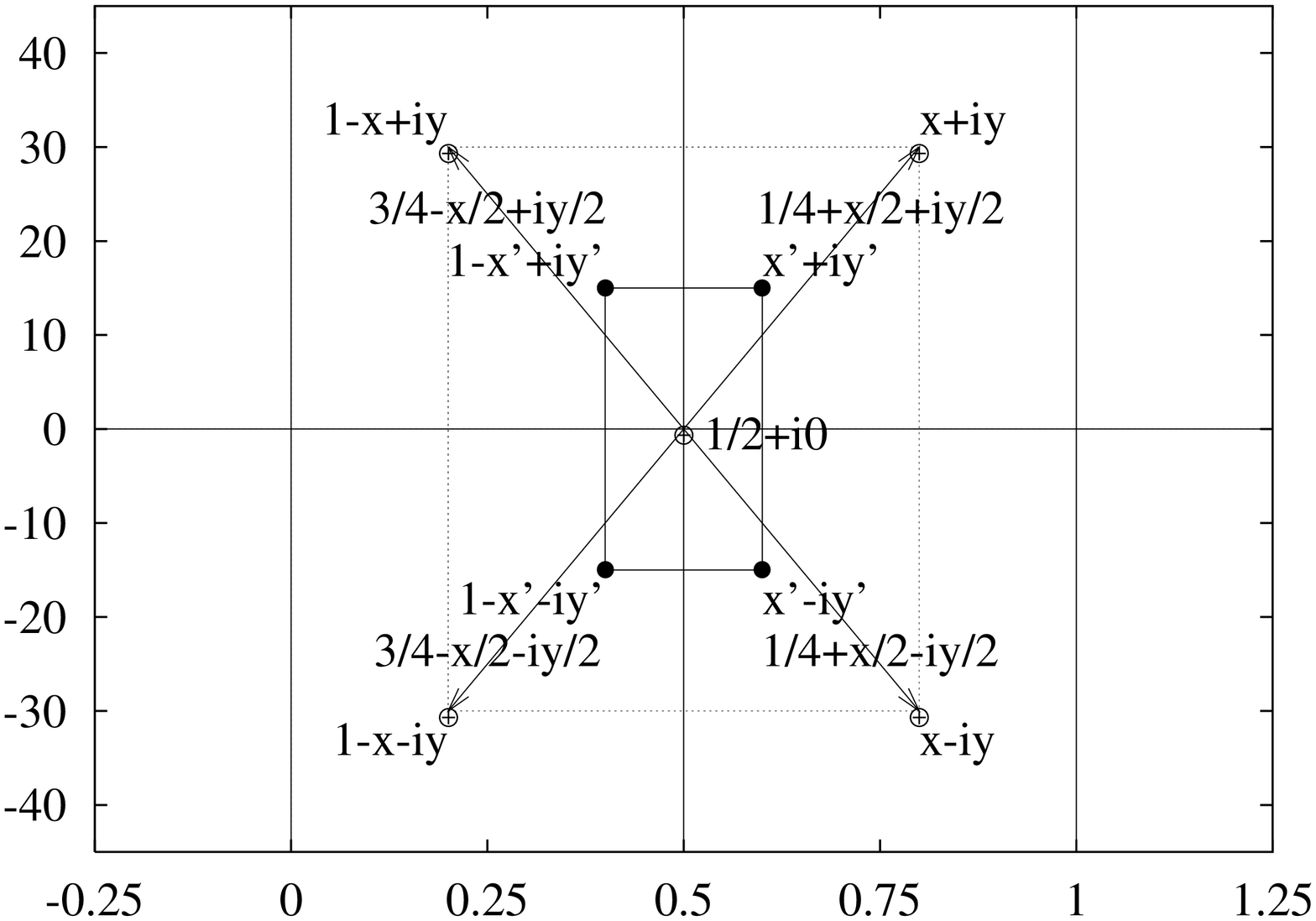,width=12cm,angle=270}}
\vspace{0.0mm}
\caption[Short Title]{The dots represent generic zeros of the $\zeta$.
The crosses represent generic states orthogonal to the reference state
$1/2+0i$. The numbers $3/4-x/2-iy/2$, etc, are the arguments of $Z$
appearing in the orthogonality relations between states orthogonal to the
reference state. Due to the functional equation of the Riemann
zeta-function (\ref{eq:RieFund}), these arguments are just the average
values between $1/2+0i$ and those orthogonal states. Here we are
referring the particular case $k=1$, $l=4$.}
\label{fig:fig}
\end{figure}

From the relation $ s' = a (s - 1/2) + 1/2 $, one can generate the
equation for a family of lines passing though the point ($x=1/2,y=0$)
given mathematically by $ (x - 1/2)/y = (x'_m - 1/2)/y'_{mn} = c_{mn} $.
``$m$'' is the label of a vertical line, and ``$n$'' the height of the
plausible zero along that vertical line. The family of lines is formed
with the diagonals of the rectangles whose vertices are the orthogonal
states (See figure \ref{fig:fig}). The zeros, for each given value of $m$,
are $s_n = x_m + i y_{mn}$. Those lines are either: (i) Parallel to the
critical Riemann line or (ii) They are not parallel. In case (i) then the
slope is infinity, so the inverse of the slope $ c_{mn}$ is zero. This
implies that $x'_m = 1/2$, the RH, if and only if, $y$ and $y'_{mn}$ are
not zero.  The family of lines $s' = as + b$, parametrized by suitable
values of ($a,b$), can be viewed as belonging to a homotopy class of maps.

When we have a continuum of the ($a,b$) or ($k,l$) parameters, we sweep a
continuum of orthogonal states located at the four vertices of a rectangle
of figure \ref{fig:fig}. The four vertices are all mapped to the four
discrete nontrivial zeros $x' + i y'$, $x' - iy'$, $(1 - x') + i y'$, $(1
- x') - i y'$, living in the four vertices of the rectangle. One can
deform continuously the rectangle containing the four orthogonal states
and shrink it in size to zero when all the four vertices collapse to the
center of symmetry ($x = 1/2, y = 0$). The center of symmetry of the
orthogonal states $s_o$ has to be mapped to the center of symmetry of the
zeros $s' = 1/2 + 0i$.

If a function is continuous in a given domain, it obeys the following, $
\lim z\to z_0\ F(z) = F(z_0) $. We will apply this result to the function
$ Z(s') = Z(as + b) = Z[a(s - 1/2) + 1/2] $. Let us imagine we have a zero
off the critical line at $ s' = x' + iy ' $, which means that the ($x,y$)
values are $ (x' - 1/2) = a (x - 1/2) $ and $ y' = a y$, $a = - 2/l$. The
argument of deforming continuously the four vertices of the rectangle,
along the diagonals, so they collapse finally to the center of symmetry $
1/2 + 0i $, after using these equations, can be expressed analytically as:
\begin{equation}
Z(s') =\lim\limits_{a\to\infty}\lim\limits_{s\to 1/2+0i} Z[a (s - 1/2) +
1/2] = \lim\limits_{a\to\infty} Z(1/2) = Z(1/2+0i).
\label{eq:limits}
\end{equation}
This is true if the function $Z$ is continuous in a given domain $0 <
{\cal R}e(s) < 1$. There is a pole of $Z (s)$ at $s = 1$ and at $s = 0$.
Therefore, when the four vertices collapse to the center point $1/2 + i0$,
one has found, after looking at the first and last term of
(\ref{eq:limits}), that $Z(s') = Z(1/2 + 0i)$. If $s' = x' + iy'$ is a
putative zero off the critical line, this would imply that $Z(s') = Z (1/2
+ 0i)$ is equal to zero, which is a contradiction since there is no zero
at $1/2 + 0i$ (an ``experimental'' fact).

Hence, we conclude that because the (homotopy) deformation does not yield
a zero at $1/2 + i0$, we cannot have zeros off the critical line at $s' =
x' + iy'$. Notice that from (\ref{eq:limits}) by using $a = - 2/l$, when
$a$ goes to $\infty$, $l$ goes to $0$, we can always approach the limits $
y = l = 0$, and $x - 1/2 = l = 0$, along straight lines in the $Y-l$ and
$X-l$ planes, respectively, whose finite slope is given by $ - y'/2 $ and
$-(x' - 1/2)/2$, respectively. For this reason the first and second terms
of (\ref{eq:limits}) are well justified. The remaining terms of
(\ref{eq:limits}) are justified based on the continuity properties of the
function $Z$. Naturaly, for fixed $a$ (not equal to $\infty$) when $s$
goes to $1/2 + 0i$ we have that $ a (s - 1/2) + 1/2 = 1/2 + 0i$ since $1/2
+ 0i$ is a fixed point.

Notice that in equation (\ref{eq:limits}) one can have $Z(s') = Z(s'')$
without having $s' = s''$ nor $s'$ equal to $1 - s''$. Of course, if this
equality holds for all values of $s'$ and $s''$, then one must have that
$s' = 1 - s''$ due to the fundamental identity $Z(s) = Z(1-s)$.

Because there is no zero at $s' = 1/2 + 0i$, this means that the state
associated with $s_o = 1/2 + 0i$ is not orthogonal to itself. So
dilating-back the point $s_o = 1/2 + 0i$, which is not an orthogonal
state, into a continuum of vertices of a rectangle, we then conclude that
the initial four vertices cannot be orthogonal states to the vacuum
$\psi_{s_o}$ and since these sates were mapped to the points $x' + i y'$,
$x' - iy'$, $(1- x') + i y'$, $(1-x') - i y'$, we conclude that these
cannot be zeros.

It is impossible to deform continuously the slopes of the diagonals of the
rectangles (given by the zeros) because this will be tantamount of saying
that the zeros (the slopes) can vary continuously. The slopes can only
change in discrete jumps if, and only if, the zeros are discrete. Then, we
cannot continuously deform the rectangle shown in figure \ref{fig:fig}
into rectangles of narrower and narrower width, and of increasing height,
which collapse to the vertical critical Riemann line in the limit of zero
width and infinite height. For this reason we can have zeros at the
critical vertical line without having a zero at the center of symmetry $s'
= 1/2 + 0i$. Roughly speaking, the diagonals and the center point $1/2 +
0i$ belong to different homotopy classes than the critical vertical line.

In order to deform the center of symmetry $ s_o = 1/2 + 0i $ onto points
belonging to the critical vertical line, it is necessary to have a family
of rhombuses with two of its vertices located in the critical vertical
line, and the other two vertices to lie in the (positive) horizontal axis.
These four vertices correspond to putative orthogonal states to the vacuum
$ \psi_{1/2 + 0i} $ and must have a one-to-one correspondence to four
putative zeros.

If this occurs, then it is possible to deform (continuously) the family of
rhombuses by varying the $a$ parameter (the $l$) continuously, in such a
way that rhombuses will then collapse into the center symmetry point $1/2
+ 0i$.

However, to begin with, this procedure is not possible for the simple
reason that there are no zeros located in the positive horizontal axis
and, hence, there is no such family of rhombuses, in the first place,
available to be deformed.  For this reason, it is not possible to deform
the center of symmetry $ 1/2 + 0i $ onto points belonging to the critical
vertical line.

And from this result, that there is no equivalence-deformation among the
points along the critical line and the center of symmetry $ 1/2 + 0i $, we
conclude that the fact that nontrivial zeros exist at the critical
vertical line is indeed compatible with the fact that there is no zero at
$ 1/2 + 0i $.

Whereas, by deforming the four vertices of the rectangles, along the
diagonals, into the center symmetry $ 1/2 + 0i $, implies that the state
$\psi_{1/2 + 0i} $ would have been orthogonal to itself, and a zero at $
1/2 + 0i $ would have been found. Since there is no zero at $ 1/2 + 0i $,
this implies that the four states living in the four vertices of the
rectangles could have not been orthogonal to the vacuum, in the first
place, and consequently, there cannot be zeros located off the critical
Riemann line.

For this reason, the core of our proof relies on symmetry considerations
and the fact that we have a continuum of differential operators that
permits us to vary the location of the orthogonal states continuously
along straight lines (the diagonals of the rectangle in figure
\ref{fig:fig}) passing through the center of symmetry.

\section{\bf A study of the symmetry of the orthogonal states to the
``vacuum'' state}
\label{sec:symmetry}

To complete the final steps of the proof of the RH we show that one can
trade-off the symmetries $ s' $ goes to $ 1 - s' $ of the function $ Z
(s') = Z (1-s') $ with the symmetries of the ``inner'' products under $ t
\rightarrow 1/t $, which, in turn, is concatenated with the $ s
\rightarrow \beta - s$ symmetry of the ``orthogonal'' states.

Relation (\ref{eq:gauss}) will be useful to show that the set
of orthogonal states to the ``vacuum'' state has the same symmetry of the
zeroes of the Riemann's zeta function. $\omega (x)$ is defined as a
summation over positive integers only, and $G (x)$ over all integers and
zero. We can recast relation (\ref{eq:gauss}) into the following useful
form, $\omega(1/x) = -1/2 + x^{1/2}/2 +x^{1/2} \omega(x)$.

Without loss of generality, we will choose $ \beta = 1 $, so the ``inner''
products are taken w.r.t the $ \psi_{1/2 +i 0} $ vacuum, and later we will
study the most general case. Invariance of the ``inner'' product under the
inversion symmetry, $ t \rightarrow 1/t$, and adopting a standard
regularization procedure of removing the infinities, yields the well
defined finite parts:
$
\langle\psi_{1/2 + i 0} (t)\vert\psi_s (t)\rangle =
\langle\psi_{1/2 + i 0} (1/t)\vert\psi_s (1/t)\rangle =
\langle\psi_{1/2 + i 0} (t)\vert\psi_{1 - s} (t)\rangle =
-Z(s') = -Z(s'')
$.
If this invariance under inversion holds for all values of $ s $ and due
to the fact that $ s' \not= s'' $ (except for the trivial case when $ 1 -s
= s $, $ s = 1/2 $) the only consistent solution, for all values of $ s $,
has to be $ s'' = 1 - s' $ due to Riemann's fundamental identity $ Z (s')
= Z (1 -s') $.

Hence ``orthogonality'' corresponds to finding a zero of zeta $ Z(s') = 0
$ inside the critical domain. The $(a,b)$ parameters are defined in terms
of $(k,l)$ in the way mentioned above, $a(k,l) = - 2/l$ and $b(k,l) = (4k
- 1)/l$, which satisfy the condition $ a + 2b = 1 $ equivalent to $ 8k - 4
= l $.

Notice that the $ t \rightarrow 1/t $ and $ s' \rightarrow 1-s' $
transformations are not modular; {\it i.e.\/} elements of the $ SL(2,R)  
$, $ SL(2,C)$ groups, respectively, because they fail to obey the
essential unit determinant condition. The Gauss-Jacobi relation is
essential to single out uniquely the transformation $ t $ goes to $ 1/t $
over all others. Only such inversions allow
\begin{equation}
\psi_s(1/t) = \psi_{1-s}(t),
\label{eq:symmpsi}
\end{equation}
The physical meaning is clear, $ t \rightarrow 1/t $ is replacing an
scaling with a contraction, it is the scaling analog of a ``time reversal
transformation'' which is translated as the conjugation $ s' $ goes to $
1-s' $. This is the key to this proof of the RH.

From the mere definition of $ \psi_s(t) $ and the Jacobi-Gauss relations,
one requires, to obey the equality $ \psi_s (1/t) = \psi_{f(s)} (t) $,
\begin{equation}
t^{s - k} e^{V(1/t)} = t^{s - k}[t^{l/4} e^{V(t)}] =
t^{- f(s) + k} e^{V(t)},
\label{eq:functionf}
\end{equation}
for all values of $ s$. The solutions to the functions $ f (s) $ are $ f
(s) = \beta - s $ where $ \beta $ is real because (\ref{eq:functionf})  
and the Gauss-Jacobi relations yield $ 2 k - \beta = l/4 $, since $ (k,l)$
are real by definition.

And from these conditions it follows that the orthogonal states to the
``vacuum'' state must obey the same symmetries properties as the zeros of
zeta, symmetric with respect to the critical line for the particular case
$ \beta = 1 $, $ s_o = 1/2 + i 0 $. If $ \psi_s $ is orthogonal to the
``vacuum'', so must be $ \psi_{1 - s} $ and by complex conjugation $
\psi_{s^*} $, $ \psi_{1 - s^*} $ due to the analyticity property of the $
Z(s) $. To show this is straightforward.

This can only occur if, and only if, $ a + 2b = 1 $. Then, the RH will
follows immediately, since,
\begin{equation}
s ' = as + b = as + {1 - a \over 2} =
a\left(s - {1 \over 2}\right) + {1\over2}.
\end{equation}

The real values of this equation are
\begin{equation}
x' = a \left(x - {1\over2}\right) + {1\over 2},
\end{equation}
and one once again, we would have arrived at the solution $ x = x' = 1/2 $
using the same arguments of the previous sections.

That ``vacuum'' state can be defined in many ways. We can show that any
``vacuum'' state must have the form $s_o = \beta/2 + i 0$. If $f(s) =
\beta - s$, the fixed point of $f$ is such that $ \beta - s_o = s_o$,
which gives $s_o = \beta/2$. We have proven that $\beta$ is a real. The
orthogonal states to the new ``vacuum'' are such that,
\begin{equation}
\langle s_o\vert s\rangle = \langle 1/2 + i 0\vert s + s_o - 1/2\rangle
= -Z [a (s + s_o - 1/2) + b] = -Z (s') = 0.
\end{equation}

If now we have,
\begin{equation}
\psi_s (1/t) = \psi_{f(s)}(t) = \psi_{\beta - s} (t),
\end{equation}
we get $ 2k - \beta = l/4 $ that can be recast in terms of $(a,b)$ as,
\begin{equation}
1 = a (2\beta - 1) + 2 b.
\label{eq:ab2beta}
\end{equation}

Invariance of the integrals under $t$ goes to $1/t$, and also of the new
``vacuum'' $s_o$, requires that we must take the inner products as,
\begin{equation}
\begin{array}{rcccl}
\langle\psi_{s_o}(1/t)\vert\psi_s(1/t)\rangle &=&
\langle\psi_{s_o}(t)\vert\psi_{\beta - s}(t)\rangle &=&
\langle\psi_{1/2}(t)\vert\psi_{\beta - s + s_o - 1/2}(t)\rangle=
\\
Z[a(\beta - s + s_o - 1/2) + b] &=& Z(1 - s') &=& 0.
\end{array}
\end{equation}
After equating the arguments inside the $Z$'s we get:
\begin{equation}
a \left(\beta - s + s_o -{1\over2}\right) + b = 1 - s',
\end{equation}
and
\begin{equation}
a \left(s + s_o - {1\over2}\right) + b = s'.
\end{equation}
One may notice that one can eliminate both $s$ and $s'$ from the last two
equations, giving the desired relation among the parameters. One gets,
after writing the new ``vacuum'' as $s_o = \beta/2$, the relation
$
1 = a (2\beta - 1) + 2 b
$,
which precisely agrees with (\ref{eq:ab2beta}) as it should, this means
that the regularization procedure was compatible with the symmetries of
the problem.

Now the zeros are given by,
\begin{equation}
s' = a \left(s + s_o - \frac{1}{2}\right) + b =
a \left(s + s_o - \frac{1}{2}\right)  + \frac{1}{2} [1 - a (2 \beta - 1)] 
=
a \left(s - \frac{\beta}{2}\right) + \frac{1}{2}.
\end{equation}
 So the real parts are,
\begin{equation}
x' = a \left(x - \frac{\beta}{2}\right) + \frac{1}{2}.
\label{eq:xxprime}
\end{equation}
This means that the $x$-value, $ \beta/2 $, obtained in
(\ref{eq:xxprime}), has to agree with the value of the $ x_o = \beta/2 $.
A discrete number of zeros requires the only solution $x = \beta/2$, which
gives again the RH result $x' = 1/2$.

Since the zeros lie in the critical line, and since the orthogonal states
with respect the new ``vacuum'' $s_o = \beta/2 = x_o + 0i$ obey the same
symmetries as the zeros, this means that the $x$-value $\beta/2$ obtained
in (\ref{eq:xxprime}) has to agree with the $x_o = \beta/2 = s_o$. And we
have seen that it does agree.

We could ask for the most general solutions of $ g (t) $ and $ f (s) $
given by the fundamental equation $ \psi_s [g (t)] = \psi_{f(s)} (t)$. The
GJ relation told us that a ``trivial'' solution is $ t $ goes to $ g (t) =
1/t$ and $ s $ goes to $ f (s) = \beta - s$, which is the origin of the
crucial correlation with $ s' $ goes to $ 1-s' $, and the clue to prove
the RH. Let us suppose that there are other solutions, $ g (t) $ is not
equal to $ 1/t $; and $ f (s) $ is not equal to $ \beta - s $. This would
imply that the new vacuum is given by the nontrivial solutions to $ f
(s_o) = s_o $ and that the new correlations are, $ t $ goes to $ g (t) $
(not equal to $ 1/t $) correlated to $ s $ goes to $ f (s) $ (not equal to
$ \beta - s $)  which is correlated to $ s' $ goes to $ 1-s' $. If these
new solutions/correlations existed, then the RH will more likely to be
false since the symmetries of the orthogonal states would have been
different.

Suppose we have found other solutions to $ \psi_s [g (t)] = \psi_{f (s)}
(t) $ than the generic (i) $ t $ goes to $ g (t) = 1/t $, (ii) $ s $ goes
to $ f(s) = \beta - s $ and (iii) the vacuum $ s_o = \beta/2 $ a fixed
point of $ f(s_o) = s_o $. If this were the case then the mapping among
orthogonal states $ s $ and zeros $ s' $ is then given by $ a (s + s^*_o -
1/2) + b = s' $ and $ a [ f(s) + s^*_o - 1/2] + b = 1 -s' $, where $ s_o $
would have been the new vacuum $ f(s_o) = s_o $. Since $ f(s) $ is assumed
different from $ \beta - s $, this means that $ s_o $ is no longer equal
to $ \beta/2$ real. The value of $ s_o $ could have been complex, so $
s_o^* $ is not equal to $ s_o $. How is it possible to eliminate
simultaneously $ s $ and $ s' $ from those two equations unless we have
the generic solutions $ f(s) = \beta - s $ linear in $ s $, that is
correlated with $ t $ goes to $ g(t) = 1/t $ and which in turn is
correlated with $ s' $ goes to $ 1-s' $?. If we cannot eliminate
simultaneously ($ s $, $ s' $) from those equations one will have a
constraint among the real parameters ($ a $, $ b $) and the complex
variables ($ s $, $ s' $), this is clearly unacceptable.

To finalize, we will see also that this symmetry of the ``vacuum'', in
the paticular case $\beta = 1 $, is also compatible with the
isospectral property of the two partner Hamiltonians,
\begin{equation}
H_A = D_2 D_1 = \left[{d\over d\ln t} - {d V(1/t)\over d\ln (1/t)} +
k\right] \left[- {d\over d\ln t} + {d V(t)\over d\ln t} + k\right],
\label{eq:partnerA}
\end{equation}
and
\begin{equation}
H_B = D_1 D_2 = \left[-{d\over d\ln t} + {d V(t)\over d\ln t} +
k\right] \left[{d \over d\ln t} - {d V(1/t)\over d\ln (1/t)} + k\right].
\label{eq:partnerB}
\end{equation}

Notice that $ V(1/t) \not= V(t)$ and for this reason $ D_2 $ is not the
``adjoint'' of $ D_1$. Operators defined on the half line do not admit an
adjoint extension, in general. Hence, the partner Hamiltonians $ H_A $, $
H_B $ are not (self-adjoint) Hermitian operators like it occurs in the
construction of SUSY QM. Consequently their eigenvalues are not real in
general.

Nevertheless one can show by inspection that if, and only if, $
\psi_s(1/t) = \psi_{1-s}(t) $ then both partner Hamiltonians are
isospectral (like in SUSY QM) whose spectrum is given by $ s(1-s) $ and
the corresponding eigenfunctions are,
\begin{equation}
H_A \psi_s(t) = s(1-s) \psi_s(t).\quad H_B \psi_s(1/t) = s
(1-s) \psi_s(1/t).
\end{equation}
Firstly by a direct evaluation one can verify,
\begin{equation}
D_1 \psi_s(t) = s \psi_s(t)\ {\rm and}\
D_2 \psi_s(1/t) = s \psi_s(1/t),
\end{equation}
{\it i.e.\/} $ \psi_s(t) $ and $ \psi_s(1/t) $ are eigenfunctions of the $
D_1 $ and $ D_2 $ operators respectively with complex eigenvalue $ s $.
Secondly, if, and only if, the condition $ \psi_s (1/t) = \psi_{1 - s} (t)
$ is satisfied then it follows that:
\begin{equation}
\begin{array}{ccccccc}
&H_B \psi_s(1/t) &=& D_1 D_2 \psi_s(1/t) &=& s D_1 \psi_s(1/t) &=\\
& s D_1 \psi_{1-s}(t) &=& s(1-s) \psi_{1-s}(t) &=& s(1-s) \psi_s(1/t),&
\end{array}
\end{equation}
meaning that $ \psi_s(1/t) $ is an eigenfunction of $ H_B $ with $ s(1 -
s)$ eigenvalue.
\begin{equation}
\begin{array}{ccccccc}
&H_A \psi_s(t) &=& D_2 D_1 \psi_s(t) &=& s D_2 \psi_s(t) &=\\
& s D_2 \psi_{1-s}(1/ t) &=& s(1-s) \psi_{1-s}(1/t) &=& s(1-s)
\psi_s(t),&
\end{array}
\end{equation}
meaning that $ \psi_s(t) $ is an eigenfunction of $ H_A $ with $s(1- s)$
eigenvalue.

Therefore, under condition $ \psi_s (1/t) = \psi_{1 - s} (t) $ the
non-Hermitian partner Hamiltonians are isospectral. The spectrum is $
s(1-s) $. Notice the similarity of these results with the eigenvalues of
the Laplace Beltrami operator in the hyperbolic plane associated with the
chaotic billiard living on a surface of constant negative curvature. In
that case the Selberg zeta function played a crucial role \cite{berry}.

The operators $ H_A $ and $ H_B$ are quadratic in derivatives like the
Laplace-Beltrami operator and involve two generalized dilatation operators
$ D_1 $ and $ D_2 $. Notice also that on the critical Riemann line, $
{\cal R}e(s) = 1/2 $, the eigenvalues are real since $ s(1-s) = s s^* $ is
real.

To sum up, the inversion properties under $ t \rightarrow 1/t$ of the
eigenfunctions of the infinite family of differential operators, $
D_1^{(k,l)}(t) $ and $ D_2^{(k,l)}(1/t) $, compatible with the existence
of an invariant ``vacuum'', are responsible for the isospectral
condition of the partner non-Hermitian Hamiltonians, $ H_A $ and $ H_B$,
like it occurs in SUSY QM. For details about the quantum inverse
scattering problem associated with the SUSY QM model which yields the
imaginary parts of the nontrivial zeros consistent with the Hilbert-Polya
proposal to prove the RH (see \cite{cc,acj}). The supersymmetric ground
state was precisely that associated with $ s_o = 1/2 + i 0 $.

\section{\bf Concluding remarks}
\label{sec:conclu}

In previous works \cite{cc,cj,acj} we have already explored a strategy
which could lead to a solution of the problem, following the Hilbert-Polya
idea. There we proposed a supersymmetric potential expressed like a
$p$-adic product. A numerical exploration of this possibility was recently
done by Wu and Sprung \cite{wusprung}. They found that the imaginary parts
of the nontrivial Riemann zeros can be reproduced using a one-dimensional
local-potential model, and that a close look at the potential suggests
that it has a fractal structure of dimension d = 1.5.

The potential found by \cite{wusprung} has a smooth part and a random
part. We believe that the fluctuating part of the potential may be
determined by using an infinite product of Weierstraas ``devil'' fractal
functions, continuous but nowhere differentiable. Also we have some
reasons to expect that more precise determination of the fractal potential
may yield one whose fractal dimension is related to Golden Mean $\phi$,
like $ d = 1 + \phi = 1.618...$

All those nice properties are in fact corollaries of the RH, if it is
proven in another way, like the one proposed in the present paper.

The ``vacuum'' state can be defined in many ways. We can show that any
``vacuum'' state must have the form $s_o = \beta/2 + i 0$. If $f(s) =
\beta - s$, the fixed point of $f$ is such that $ \beta - s_o = s_o$,
which gives $s_o = \beta/2$, where $ \beta $ was shown to be real.

Hence the orthogonal sates to the ``vacuum'' $s_o = \beta/2 + i 0$ are
reflection symmetric with respect to the point $s_o$, in the same way that
the zeros of zeta must be reflection symmetric with respect to the point
$1/2 + i 0$. Of course, we must always include the complex conjugates.

We found that irrespective of the choice of $\beta $ we always get $ s' =
a (s - \beta/2) + 1/2 $, whose reals parts are $x' = a (x - \beta/2) +
1/2$. If the zeros are discrete the only solution is $x = \beta/2$, which
means $x'= 1/2$, so the RH is true.

For consistency purposes, since the zeros collapse to one line, the value
$x = \beta/2$ must agree with the center of the reflection symmetry, with
the value of $s_o = \beta/2$.

In this way, for any $\beta$, for all $f (s) = \beta - s$, and for $\psi_s
(1/t) = \psi_{\beta -s} (t)$ we have found fully consistent results that
yield $x' = 1/2$ always. The RH is true.

What we have shown is that there is a concatenation among the three
transformations: (i) $ t$ goes to $ 1/t $. (ii) $ s $ goes to $ \beta - s
$, where $\beta$ has to be real because of the condition derived from the
Gauss-Jacobi identity in $\psi_s (1/t) = \psi_{(\beta - s)} (t)$ yields
that $2k - \beta = l/4$, since $(k,l)$ are real like $t$, this forces
$\beta$ to be real. (iii) $ s' $ goes to $ 1- s' $.

After adopting a regularization of dropping the infinities and which is
indeed consistent with the symmetries, as we have shown, then we have that
the above concatenation of three transformations (i), (ii), (iii),
manifest itself as follows.

Let us take $\beta/2+i0$ the ``vacuum'' or invariant state. The result is
valid for all $\beta$ after being careful how one takes the ``inner''
products w.r.t the new vacuum $\beta/2$.

Hence, after dropping the infinities, for a general $\beta$, $s_o =
\beta/2$, we get the fundamental relation $ 1 = a (2 \beta - 1) + 2b $,
which leads to $ s' = a (s - \beta/2) + 1/2 $, that finally yields the
proof of the RH based on the discreteness properties of the zeros.

What ties all these identities together based on the above concatenation
of three transformations, is nothing but the fundamental identity $ Z (s')
= Z (1-s') $ and the Gauss-Jacobi relations (after regularization). Of
course, we have to use the formula of \cite{karatsuba} (after dropping the
infinities) and insert the factors of 2 due to our summation over the
integers from $-\infty$ to $+\infty$.

One further remark is in order. Even if we fix the values of $ s $, the
symmetry-invariance of the ``inner'' products under inversions $ t $ goes
to $ 1/t $ still gives $ -Z (s') = -Z (s'') $. If we fix the values of $s$
of the ``orthogonal'' states, one still has the freedom to vary in
continuous manner the $ (k,l) $ parameters, i.e the $ (a,b) $ coefficients
which appear inside the arguments of the $ Z $ functions. If the above
equation holds for all values of $ (k,l) $ (for all values of $(a,b)$),
and since $ s' $ is not equal to $ s''$ in general, then the only
possibility to obey the above equality, for all values of $ (k,l) $ or $
(a,b) $ is to have $s'' = 1 - s'$ due to the fundamental identity $Z (s')
= Z (1-s')$. And once more, for fixed values of $ s $, we still can
trade-off the $ s' $ goes for $ 1 - s' $ symmetry for the inversions $ t
\rightarrow 1/t $ symmetry.

Now we sum up that we have found. The clues to our proof of the RH are the
following.

1. The Gauss-Jacobi relation of the theta series $G (1/x) = x^{1/2} G (x)$
that requires summing over all negative, zero and positive integers.

2. To translate the fundamental symmetry $Z (s') = Z (1-s')$ as the
symmetry of $t$ goes to $1/t$.

3. Adopting a standard regularization program of dropping the infinities
and retaining the well defined terms $ -Z (s') = -Z(1-s')$ in the
integrals of the book \cite{karatsuba}.

4. For the particular case when $ \beta = 1 $, one has $s_o = 1/2 + i 0$
for the invariant ``vacuum'' state compatible with the $s$ goes to $1 - s$
and the $t$ goes to $1/t$ symmetries. In general the ``vacuum'' is given
by $ s_o = \beta/2 $, $ \beta $ real, therefore all inner products can
always be written in the form $\langle s_1\vert s_2\rangle = \langle
s_o\vert s_1^* + s_2 - s_o\rangle$ and repeat our arguments all over again
by defining $s = s_1^* + s_2 - s_o$.

5. By deforming the four vertices of the rectangles, along the diagonals,
into the center symmetry $ 1/2 + 0i $, implies that the state $\psi_{1/2 +
0i} $ would have been orthogonal to itself, and a zero at $ 1/2 + 0i $
would have been found. Since there is no zero at $ 1/2 + 0i $, this
implies that the four states living in the four vertices of the rectangles
could have not been orthogonal to the vacuum, in the first place, and
consequently, there cannot be zeros located off the critical Riemann line.

\section*{\bf Acknowledgements}

We thank the Center for Theoretical Studies of Physical Systems, Clark
Atlanta University, Atlanta, Georgia, USA, and the Research Committee of
the University of Antioquia (CODI), Medell\'{\i}n, Colombia for support.
Very constructive comments from Matthew Watkins, School of Mathematical
Sciences University of Exeter, Gottfried Curio from Institute of Physics,
Humboldt University, and Emilio Elizalde from Universitat de Barcelona are
gratefully acknowledged. CC is indebted to the Perelman and Kuhlmann
family for their very kind hospitality in New York and Heidelberg,
Germany, where this work was completed.

\newpage

\section*{\bf Figures}
\label{sec:figs}
 
\def\baselinestretch{1}

\vfill

\end{document}